\documentclass[aip,apl,reprint,superscriptaddress,floatfix]{revtex4-1}
\usepackage{graphicx}
\usepackage[ansinew]{inputenc}
\usepackage{array}
\usepackage{color}
\usepackage{amsmath}
\usepackage{amsxtra}
\usepackage{amstext}
\usepackage{amssymb}
\usepackage{latexsym}

\begin{document}

\title{Focal overlap gating in velocity map imaging to achieve high signal-to-noise ratio in photo-ion pump-probe experiments}

\author{Niranjan Shivaram}
\email{nhshivaram@lbl.gov}
\affiliation{Chemical Sciences Division, Lawrence Berkeley National Laboratory, Berkeley, CA 94720}

\author{Elio Champenois}
\affiliation{Chemical Sciences Division, Lawrence Berkeley National Laboratory, Berkeley, CA 94720}
\affiliation{Graduate Group in Applied Science and Technology, University of California, Berkeley, CA 94720}

\author{James Cryan}
\affiliation{PULSE Institute for Ultrafast Energy Science, SLAC National Accelerator Laboratory, Menlo Park, CA 94025}

\author{Travis Wright}
\affiliation{Chemical Sciences Division, Lawrence Berkeley National Laboratory, Berkeley, CA 94720}
\affiliation{Department of Chemistry, University of California at Davis, Davis, CA 95616}

\author{Taylor Wingard}
\affiliation{Chemical Sciences Division, Lawrence Berkeley National Laboratory, Berkeley, CA 94720}
\affiliation{Department of Physics, University of California, Berkeley, CA 94720}

\author{Ali Belkacem}
\affiliation{Chemical Sciences Division, Lawrence Berkeley National Laboratory, Berkeley, CA 94720}

\date{\today}

\begin{abstract}

We demonstrate a new technique in velocity map imaging (VMI) that allows spatial gating of the laser focal overlap region in time resolved pump-probe experiments. This significantly enhances signal-to-noise ratio by eliminating background signal arising outside the region of spatial overlap of pump and probe beams. This enhancement is achieved by tilting the laser beams with respect to the surface of the VMI electrodes which creates a gradient in flight time for particles born at different points along the beam. By suitably pulsing our microchannel plate detector, we can select particles born only where the laser beams overlap. This spatial gating in velocity map imaging can benefit nearly all photoion pump-probe VMI experiments especially when extreme-ultraviolet (XUV) light or X-rays are involved which produce large background signals on their own.

\end{abstract}

%\keywords{}

\maketitle

Velocity Map Imaging (VMI) \cite{Eppink1997, Parker1997} is a technique for imaging charged particles which is now widely used in atomic, molecular and chemical physics experiments to study a variety of different processes in gas phase systems \cite{}. It has become a standard technique to measure ultrafast processes in these systems on attosecond and femtosecond time scales using vacuum-ultraviolet (VUV) and extreme-ultraviolet (XUV) light from high-order harmonic generation (HHG) and Free Electron Laser (FEL) sources\cite{sansone2010, trabattoni2015, Champenois2016, sato2016, johnsson2008}, and to study X-ray driven processes from synchrotron sources \cite{o2011}. VMI can provide angle and energy resolved photo-ion and photo-electron spectra in a simple, economical setup without the requirement for extensive post acquisition analysis except an inversion procedure that can be achieved in a few different ways \cite{dribinski2002, manzhos2003, garcia2004, roberts2009}. Since the time it was first introduced, VMI has evolved and has been modified and improved, for example, to obtain three dimensional velocity distributions using time-slicing \cite{gebhardt2001, townsend2003, lee2014communication}, to image ions and electrons in coincidence \cite{rolles2007, lee2014} for a complete reconstruction of a photon-molecule interaction event, etc. 

\begin{figure}
\includegraphics[width=0.4 \textwidth]{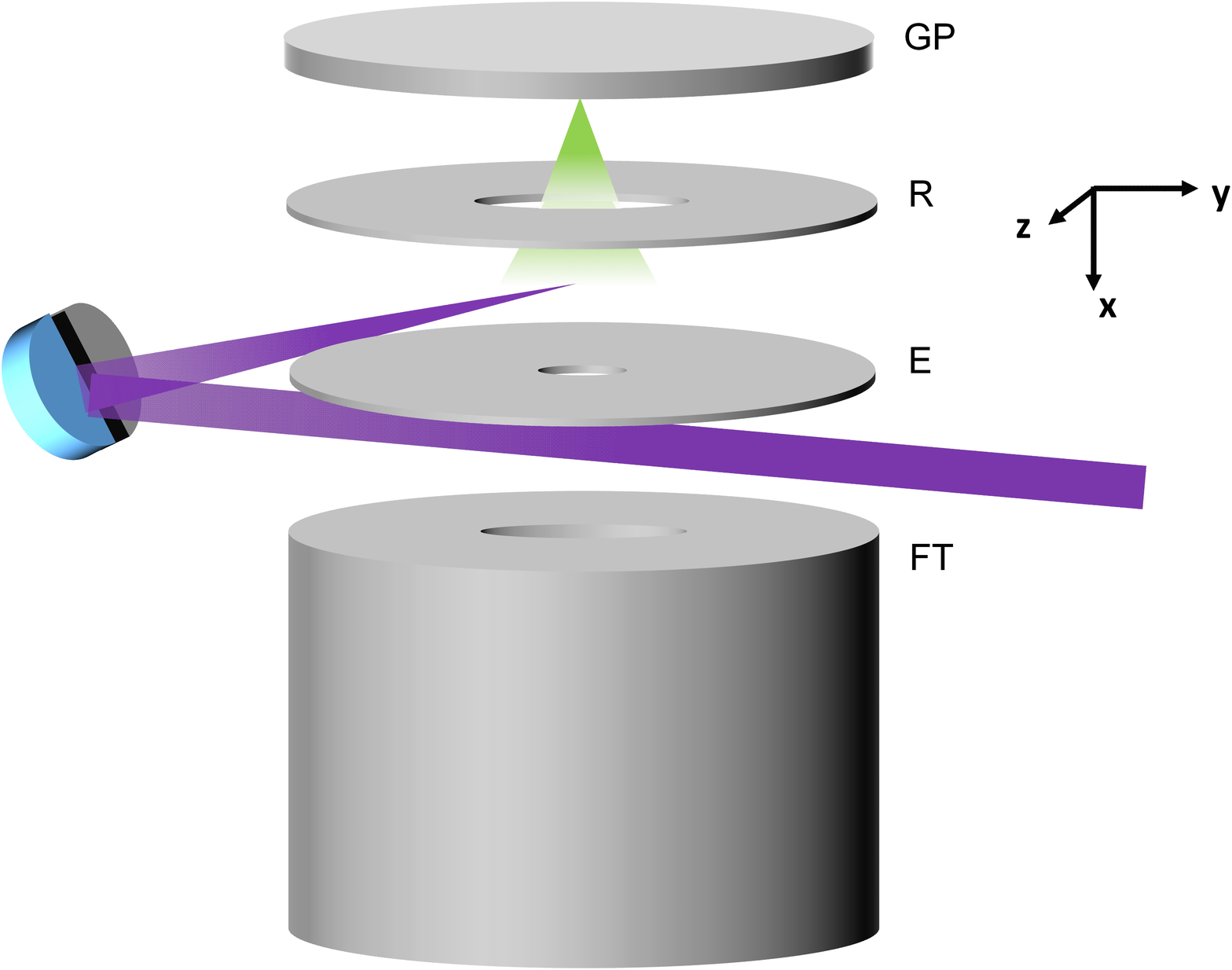}
\caption{\label{fig:1} A schematic of the velocity map imaging (VMI) setup. The laser beam is split into pump and probe and back focused at a tilted angle by a curved split mirror. The target gas is injected from a 1 micron hole in the gas plate (GP). R - Repeller, E - Extractor, FT - Flight Tube. A suitable voltage applied on a mesh - M (not shown) located at the end of the flight tube allows us to reject any ions originating from the unfocused beam between the extractor and flight tube. A micro-channel plate (MCP) detector (not shown) images the charged particles.}
\end{figure}

Here, we present a new technique which involves a simple modification of the input laser beam geometry in a standard VMI which results in significant enhancement of signal-to-noise ratio in time resolved pump-probe measurements. In typical pump-probe experiments using VMI, the pump and probe laser beams are focused on a gas target at the center of the VMI setup where they are temporally and spatially overlapped. The laser beams propagate along a line parallel to the VMI electrodes and the charged particles born along this line are imaged by the VMI spectrometer. Since these particles are born at the same distance from the repeller electrode, their time-of-flight to the detector for a given mass is the same irrespective of distance along the laser propagation direction. In our method, we make a simple modification of this standard beam geometry by introducing a tilt in the laser beams propagating through the VMI setup (see Fig. \ref{fig:1}). This is effectively a `passive streaking' of the particles which creates a variation in time-of-flight for particles born at different points along the laser beam. By appropriately pulsing our microchannel plate detector to select only particles born in the overlap region of the laser beams, it is possible to reduce and even eliminate background signal arising from regions outside the overlap zone. The only trade-off with this technique is a possible energy smearing of the particles which could introduce uncertainties in the energy measurement. We show that this smearing is almost negligible for the parameters in our setup.

The experimental setup is described elsewhere \cite{Champenois2016}. Briefly, it consists of a three stage Ti:Sapphire laser amplifier which can provide 20 mJ, 25 fs near infra-red (IR) laser pulses at 780 nm central wavelength with a repetition rate of 1 kHz. This laser beam is focused into a cell containing Ar gas where we generate high harmonic VUV and XUV radiation by the process of high order harmonic generation \cite{macklin1993, krause1992, corkum1993}. These high harmonics are routed through grazing incidence mirrors which also reject the residual IR light. The harmonics are then focused onto a target gas (ethylene in this measurement) in the center of the VMI spectrometer using split back focusing mirrors at an angle of 14 degrees with the horizontal (Fig. \ref{fig:1}). The moveable split mirror allows us to split the laser beam to produce pump and probe pulses which can be time delayed relative to each other by upto 3 ps. The resolution in time delay is less than 1 fs. The target gas is injected to the center of the VMI through a 1 micron diameter hole in a plate above the repeller electrode. In our design, the repeller electrode has a 25 mm diameter hole in the center. The relevant voltages applied to the VMI electrodes are shown in table \ref{tab:1}. The ions produced by the laser are accelerated towards a microchannel plate detector coupled to a phosphor screen which is then imaged by CMOS camera.

\begin{table}
%\resizebox{0.48 \textwidth}{!}
\begin{center}
\begin{tabular}{l *{5}{c}}
                           & GP (V)        & R (V)        & E (V)         & FT (V)     & M (V) \\
\hline							
SI Mode                    & 2500          & 3000         & 180           & 0          &  200   \\
VMI Mode                   & 2500          & 3000         & 1930          & 0          &  2000   \\
\hline  
\end{tabular}
\end{center}
\caption{\label{tab:1} The voltage settings for different electrodes in the Spatial Imaging (SI) mode and Velocity Map Imaging (VMI) mode. GP - Gas Plate, R - Repeller, E - Extractor, FT - Flight Tube and M - Mesh. The distance between the electrode plates is 20 mm. The diameter of the holes in the repeller, extractor and flight tube are respectively 25.4 mm, 7 mm and 8.4 mm. The MCP detector is located 850 mm from the interaction region.}
\end{table}

\begin{figure}[t]
\begin{center}
\includegraphics[width=0.5 \textwidth]{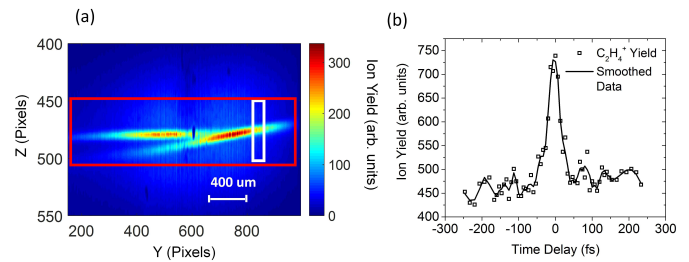}
\caption{\label{fig:2} (a) A spatial mode image of C$_2$H$_4^+$ ions. This ion image provides a direct microscopic image of the focal region of the pump and probe arms. The white and the red box represent `image gates' of two different sizes. The magnification under the voltage conditions used is $\sim$ 30. The actual spatial scale in the focal region is also shown. (b) Pump-probe delay scan of C$_2$H$_4^+$ obtained in spatial imaging mode. The yield is obtained by integrating ion yield over the white image gate (box) shown in (a). A solid line representing a 5-point smoothing of the data is shown as a guide to the eye.}
\end{center}
\end{figure}

We use ethylene (C$_2$H$_4$) as a target gas and use harmonics consisting predominantly of the third harmonic (260 nm, 4.75 eV) in both pump and probe arms to excite and ionize ethylene. We scan the time delay between pump and probe arms and record the C$_2$H$_4^+$ yield which shows a cross correlation peak when the two pulses are overlapped in time. We first record this yield in a spatial imaging mode which allows a one-to-one mapping of point of birth to a given point on the detector thus providing us a direct image of the focal region \cite{Shivaram2010}. This and other modes of imaging with a VMI setup are discussed in detail in \cite{Stei2013}. Imaging in this mode allows us to clearly isolate the signal originating only from the pump-probe overlap region and gives us an upper limit on the achievable contrast of the transient effect in a pump-probe delay scan. Figure \ref{fig:2} (a) shows an image of the focal region obtained in the spatial imaging mode for C$_2$H$_4^+$ ions which have less than 20 meV kinetic energy. It is important to note that the spatial imaging mode to measure time delay dependent ion yields only works well for very low kinetic energy ($\sim$ 10 - 20 meV in our electric field conditions). For higher kinetic energies, the image becomes blurred and results in a broad stripe instead of a sharp line. Also the energy and angular distribution of the ions cannot be measured in this mode. 

In contrast to the spatial imaging mode, VMI mode provides energy and angular distributions of the charged particles irrespective of their point of origin. This is a powerful imaging mode that is widely used to obtain momentum distributions of ions and electrons especially in time-resolved pump-probe experiments. One of the main requirements of a pump-probe measurement is that only the signal originating from the pump and probe overlap region be recorded. The signal from pump and probe on their own constitutes an undesirable background. This is especially a problem when XUV or X-rays are involved since they inevitably ionize all target atoms or molecules. Since the VMI mode `averages' over all source points, the time dependent pump-probe signal gets buried in the background. Typically, this problem is mitigated by using a confined supersonic gas jet as the target. The gas jet is positioned such that the gas density is highest in the pump-probe overlap region and very small elsewhere. The size of the gas jet is usually on the order of 0.5 - 1 mm and such a confined target can enhance signal-to-noise contrast. Our technique presented here is a simple alternate approach to increase signal contrast in VMI mode pump-probe photo-ion spectroscopy experiments without the need for a highly confined gas jet. Even in situations where a cold supersonic jet is essential, this technique may further improve signal contrast.

\begin{figure}[t]
\begin{center}
\includegraphics[width=0.48 \textwidth]{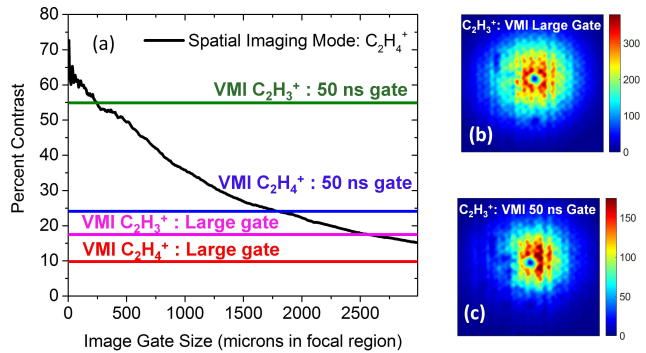}
\caption{\label{fig:3} (a) Percent contrast of the time delay dependent pump-probe signal. The contrast for C$_2$H$_4^+$ in spatial imaging mode for different image gate sizes (black line) increases with decreasing gate size as expected. The horizontal colored lines represent contrast obtained in the VMI mode using our technique for C$_2$H$_4^+$ and C$_2$H$_3^+$ ions. A significant increase in contrast is seen when a 50 ns time gate is applied compared to a large time gate (corresponding to full focal averaging). A raw VMI detector image of C$_2$H$_3^+$ ions with a large time gate (b) and 50 ns gate (c) are shown. The hole in the center of both images and on the left side in (c) is due to MCP damage.}
\end{center}
\end{figure}

Figure \ref{fig:2} (b) shows a pump-probe delay scan of C$_2$H$_4^+$ with 260 nm pulses as both pump and probe. This data was obtained in the spatial imaging mode by gating the image (fig. \ref{fig:2} (a)) near the focal overlap region. We define the contrast for this delay scan as a ratio of the peak ion yield to the average ion yield between -200 fs and -100 fs. This contrast is found for different image gate (box) sizes and is plotted as a black line in figure \ref{fig:3} (a). As expected, the contrast value increases with decreasing box size with a maximum around 60 \% and a minimum around 15 \%. Thus, the spatial imaging mode allows us to set limits on the maximum achievable contrast for a pump-probe delay scan of C$_2$H$_4^+$ under our conditions.

We now discuss the same pump-probe measurement in the VMI mode. Here, we apply a time gate to our detector by pulsing the front plate of the MCP. Two different widths are used for the time gate - a large gate ($\sim$ 200 ns) and a 50 ns gate. The large time gate allows us to image C$_2$H$_4^+$ ions originating from the entire focal region of about 3 mm. Whereas the 50 ns wide gate allows us to select ions arising from a much smaller region near the overlap of the pump and probe beams. We find the optimal location in time for the 50 ns gate by systematically repeating the pump-probe delay scan for different positions of the time gate and measuring the resulting contrast in the delay scan. Alternatively, this optimal time location of the gate can be found from simulation (SIMION) if the exact location of the focal overlap region is known. The contrast obtained for C$_2$H$_4^+$ with the 50 ns gate and the large gate are shown by the blue and red horizontal lines respectively in figure \ref{fig:3} (a). The contrast increases from about 10 \% with the large gate (full focal averaging) to about 25 \% with the 50 ns gate. We repeat the same measurement for the C$_2$H$_3^+$ fragment in VMI mode. The contrast obtained in this case are shown by the magenta and the olive lines. In this case the increase in contrast is from around 17 \% to 55 \%. The reason for the larger increase in contrast in the case of C$_2$H$_3^+$ is the presence of other higher harmonics in both pump and probe arms that produce more C$_2$H$_3^+$ fragments compared to C$_2$H$_4^+$. Our gating technique reduces this background and increases the pump-probe signal contrast.

Figures \ref{fig:3} (b) and (c) show raw VMI detector images of C$_2$H$_3^+$ which have $\sim$ 0.1 eV kinetic energy. These images were obtained by adding individual ion hits on the MCP using a hit-finding algorithm. The holes at the center of the two images and on the left side in figure \ref{fig:3} (c) are artifacts due to MCP damage. The hole on the left is not visible in figure \ref{fig:3} (b) due to aberrations in the imaging associated with a larger spatial size of the ion source points. Though VMI mode maps the momentum of particles to points on the detector irrespective of the spatial location of the source points, for large size of the source (relative to the size of the extractor hole) the image on the detector has aberrations which create a spread in the location of the center of the image. Thus, in the case of large time gate (fig. \ref{fig:3} (b)), which has a larger spatial size of source points, the counts on the left side of the image are larger than for the short gate case (fig. \ref{fig:3} (c)) making the damage on the MCP not visible in figure \ref{fig:3} (b).

\begin{figure*}
\begin{center}
\includegraphics[width=0.90 \textwidth]{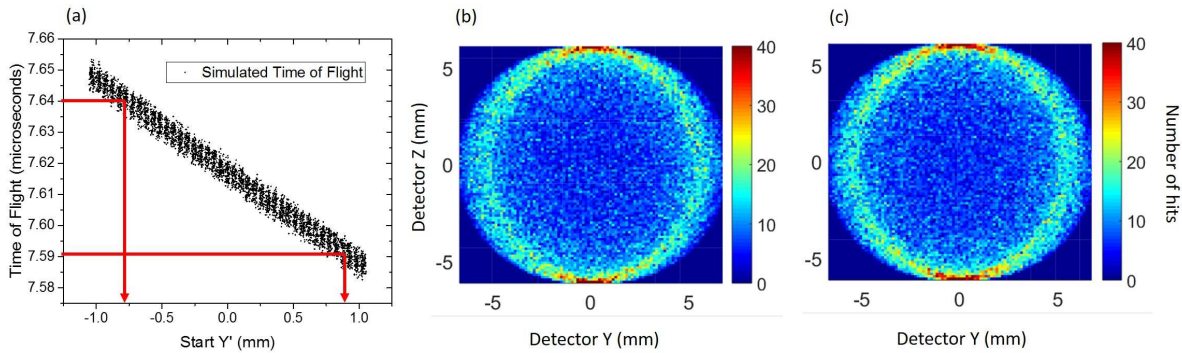}
\caption{\label{fig:4} (a) A SIMION simulation of the C$_2$H$_4^+$ ion time of flight (VMI mode) for a tilted line source as in our experiment. A gradient in flight times is seen depending on the point of origin. $Y'$ is the coordinate along the laser propagation direction (see text). A 50 ns time gate on the detector (indicated by red lines) allows us to select ions from $\sim$ 1.7 mm around the focal overlap region. (b) A histogram plot of the simulated VMI mode hits on the detector for a tilted line source and (c) a straight line source (parallel to the VMI electrodes). A tilted source introduces negligible distortions in the image.}
\end{center}
\end{figure*} 

The actual size of the focal overlap region selected by our short time gate can be obtained from SIMION simulations. Figure \ref{fig:4} (a) shows the time of flight of C$_2$H$_4^+$ ions in VMI mode as a function of the laser propagation co-ordinate $Y'$ in the focal region. $Y' = Y/cos(\theta)$ where $\theta$ is the angle of tilt of the focusing laser beam with respect to the surface of the extractor electrode. $\theta$ is 14 degrees in our experiment. The red lines in figure \ref{fig:4} (a) indicate the size of the focal region selected by a 50 ns time gate. Under our conditions, a 50 ns time gate corresponds to selecting an $\sim$ 1.7 mm region around the focal overlap. This agrees very well with the experimentally obtained value of 1.75 mm given by the point of intersection of the black curve and the blue line in figure \ref{fig:3} (a). Though this is not a very small region compared to what can be achieved using a confined gas jet, as seen from the discussion above, a significant enhancement in contrast can be obtained with relatively less effort. Using an even shorter pulse on the order of 5 ns, selectivity of a few hundred microns around the focal overlap can be achieved. A short gate can also be achieved by increasing the tilt of the beam as long as the uncertainty in energy measurement introduced by the tilt is acceptable. While we have demonstrated this technique for photo-ion measurements, with suitable modification of the VMI design and the applied voltages combined with a few nanosecond time gate and a larger tilt in the laser beam, it could potentially work for photo-electron spectroscopy as well. 

Finally, we perform SIMION simulations to investigate aberrations introduced by a tilted line source of ions compared to a line source that is parallel to the VMI electrodes. Figures \ref{fig:4} (b) and (c) show histogram plots of simulated VMI mode hits of C$_2$H$_4^+$ on the detector, with a kinetic energy of 0.1 eV, for a tilted line source and a straight line source respectively. As can be seen from these figures, the amount of aberrations introduced by the tilt in the focusing laser is negligible. 
 
In conclusion, we have developed a new technique in velocity map imaging to significantly enhance pump-probe signal contrast in photoion imaging experiments. A simple tilt in the focusing laser beams combined with a short time gate on the micro-channel plate detector is sufficient to achieve a large improvement in signal contrast. We believe this technique can be very beneficial for attosecond and femtosecond pump-probe photoion experiments involving extreme-ultraviolet and x-ray light from high-order harmonic generation/Free Electron Lasers and experiments combining lasers and x-rays at Synchrotron facilities.  

%\begin{acknowledgments}
We thank Dr. Daniel Slaughter for helpful comments and suggestions. This work was supported by the U.S. Department of Energy, Office of Science, Office of Basic Energy Sciences, Chemical Sciences, Geosciences, and Biosciences Division under Contract No. DE-AC02-05CH11231.

\bibliography{VMI_gating_refs}

\begin{thebibliography}{22}%
\makeatletter
\providecommand \@ifxundefined [1]{%
 \@ifx{#1\undefined}
}%
\providecommand \@ifnum [1]{%
 \ifnum #1\expandafter \@firstoftwo
 \else \expandafter \@secondoftwo
 \fi
}%
\providecommand \@ifx [1]{%
 \ifx #1\expandafter \@firstoftwo
 \else \expandafter \@secondoftwo
 \fi
}%
\providecommand \natexlab [1]{#1}%
\providecommand \enquote  [1]{``#1''}%
\providecommand \bibnamefont  [1]{#1}%
\providecommand \bibfnamefont [1]{#1}%
\providecommand \citenamefont [1]{#1}%
\providecommand \href@noop [0]{\@secondoftwo}%
\providecommand \href [0]{\begingroup \@sanitize@url \@href}%
\providecommand \@href[1]{\@@startlink{#1}\@@href}%
\providecommand \@@href[1]{\endgroup#1\@@endlink}%
\providecommand \@sanitize@url [0]{\catcode `\\12\catcode `\$12\catcode
  `\&12\catcode `\#12\catcode `\^12\catcode `\_12\catcode `\%12\relax}%
\providecommand \@@startlink[1]{}%
\providecommand \@@endlink[0]{}%
\providecommand \url  [0]{\begingroup\@sanitize@url \@url }%
\providecommand \@url [1]{\endgroup\@href {#1}{\urlprefix }}%
\providecommand \urlprefix  [0]{URL }%
\providecommand \Eprint [0]{\href }%
\providecommand \doibase [0]{http://dx.doi.org/}%
\providecommand \selectlanguage [0]{\@gobble}%
\providecommand \bibinfo  [0]{\@secondoftwo}%
\providecommand \bibfield  [0]{\@secondoftwo}%
\providecommand \translation [1]{[#1]}%
\providecommand \BibitemOpen [0]{}%
\providecommand \bibitemStop [0]{}%
\providecommand \bibitemNoStop [0]{.\EOS\space}%
\providecommand \EOS [0]{\spacefactor3000\relax}%
\providecommand \BibitemShut  [1]{\csname bibitem#1\endcsname}%
\let\auto@bib@innerbib\@empty
%</preamble>
\bibitem [{\citenamefont {Eppink}\ and\ \citenamefont
  {Parker}(1997)}]{Eppink1997}%
  \BibitemOpen
  \bibfield  {author} {\bibinfo {author} {\bibfnamefont {A.~T. J.~B.}\
  \bibnamefont {Eppink}}\ and\ \bibinfo {author} {\bibfnamefont {D.~H.}\
  \bibnamefont {Parker}},\ }\href {<Go to ISI>://WOS:A1997XX25700028}
  {\bibfield  {journal} {\bibinfo  {journal} {Review of Scientific
  Instruments}\ }\textbf {\bibinfo {volume} {68}},\ \bibinfo {pages} {3477}
  (\bibinfo {year} {1997})}\BibitemShut {NoStop}%
\bibitem [{\citenamefont {Parker}\ and\ \citenamefont
  {Eppink}(1997)}]{Parker1997}%
  \BibitemOpen
  \bibfield  {author} {\bibinfo {author} {\bibfnamefont {D.~H.}\ \bibnamefont
  {Parker}}\ and\ \bibinfo {author} {\bibfnamefont {A.~T. J.~B.}\ \bibnamefont
  {Eppink}},\ }\href {<Go to ISI>://WOS:A1997XQ33200022} {\bibfield  {journal}
  {\bibinfo  {journal} {Journal of Chemical Physics}\ }\textbf {\bibinfo
  {volume} {107}},\ \bibinfo {pages} {2357} (\bibinfo {year}
  {1997})}\BibitemShut {NoStop}%
\bibitem [{\citenamefont {Sansone}\ \emph {et~al.}(2010)\citenamefont
  {Sansone}, \citenamefont {Kelkensberg}, \citenamefont {P{\'e}rez-Torres},
  \citenamefont {Morales}, \citenamefont {Kling}, \citenamefont {Siu},
  \citenamefont {Ghafur}, \citenamefont {Johnsson}, \citenamefont {Swoboda},
  \citenamefont {Benedetti} \emph {et~al.}}]{sansone2010}%
  \BibitemOpen
  \bibfield  {author} {\bibinfo {author} {\bibfnamefont {G.}~\bibnamefont
  {Sansone}}, \bibinfo {author} {\bibfnamefont {F.}~\bibnamefont
  {Kelkensberg}}, \bibinfo {author} {\bibfnamefont {J.}~\bibnamefont
  {P{\'e}rez-Torres}}, \bibinfo {author} {\bibfnamefont {F.}~\bibnamefont
  {Morales}}, \bibinfo {author} {\bibfnamefont {M.~F.}\ \bibnamefont {Kling}},
  \bibinfo {author} {\bibfnamefont {W.}~\bibnamefont {Siu}}, \bibinfo {author}
  {\bibfnamefont {O.}~\bibnamefont {Ghafur}}, \bibinfo {author} {\bibfnamefont
  {P.}~\bibnamefont {Johnsson}}, \bibinfo {author} {\bibfnamefont
  {M.}~\bibnamefont {Swoboda}}, \bibinfo {author} {\bibfnamefont
  {E.}~\bibnamefont {Benedetti}},  \emph {et~al.},\ }\href@noop {} {\bibfield
  {journal} {\bibinfo  {journal} {Nature}\ }\textbf {\bibinfo {volume} {465}},\
  \bibinfo {pages} {763} (\bibinfo {year} {2010})}\BibitemShut {NoStop}%
\bibitem [{\citenamefont {Trabattoni}\ \emph {et~al.}(2015)\citenamefont
  {Trabattoni}, \citenamefont {Klinker}, \citenamefont
  {Gonz{\'a}lez-V{\'a}zquez}, \citenamefont {Liu}, \citenamefont {Sansone},
  \citenamefont {Linguerri}, \citenamefont {Hochlaf}, \citenamefont {Klei},
  \citenamefont {Vrakking}, \citenamefont {Mart{\'\i}n} \emph
  {et~al.}}]{trabattoni2015}%
  \BibitemOpen
  \bibfield  {author} {\bibinfo {author} {\bibfnamefont {A.}~\bibnamefont
  {Trabattoni}}, \bibinfo {author} {\bibfnamefont {M.}~\bibnamefont {Klinker}},
  \bibinfo {author} {\bibfnamefont {J.}~\bibnamefont
  {Gonz{\'a}lez-V{\'a}zquez}}, \bibinfo {author} {\bibfnamefont
  {C.}~\bibnamefont {Liu}}, \bibinfo {author} {\bibfnamefont {G.}~\bibnamefont
  {Sansone}}, \bibinfo {author} {\bibfnamefont {R.}~\bibnamefont {Linguerri}},
  \bibinfo {author} {\bibfnamefont {M.}~\bibnamefont {Hochlaf}}, \bibinfo
  {author} {\bibfnamefont {J.}~\bibnamefont {Klei}}, \bibinfo {author}
  {\bibfnamefont {M.}~\bibnamefont {Vrakking}}, \bibinfo {author}
  {\bibfnamefont {F.}~\bibnamefont {Mart{\'\i}n}},  \emph {et~al.},\
  }\href@noop {} {\bibfield  {journal} {\bibinfo  {journal} {Physical Review
  X}\ }\textbf {\bibinfo {volume} {5}},\ \bibinfo {pages} {041053} (\bibinfo
  {year} {2015})}\BibitemShut {NoStop}%
\bibitem [{\citenamefont {Champenois}\ \emph {et~al.}(2016)\citenamefont
  {Champenois}, \citenamefont {Shivaram}, \citenamefont {Wright}, \citenamefont
  {Yang}, \citenamefont {Belkacem},\ and\ \citenamefont
  {Cryan}}]{Champenois2016}%
  \BibitemOpen
  \bibfield  {author} {\bibinfo {author} {\bibfnamefont {E.~G.}\ \bibnamefont
  {Champenois}}, \bibinfo {author} {\bibfnamefont {N.~H.}\ \bibnamefont
  {Shivaram}}, \bibinfo {author} {\bibfnamefont {T.~W.}\ \bibnamefont
  {Wright}}, \bibinfo {author} {\bibfnamefont {C.~S.}\ \bibnamefont {Yang}},
  \bibinfo {author} {\bibfnamefont {A.}~\bibnamefont {Belkacem}}, \ and\
  \bibinfo {author} {\bibfnamefont {J.~P.}\ \bibnamefont {Cryan}},\ }\href {<Go
  to ISI>://WOS:000368617300009} {\bibfield  {journal} {\bibinfo  {journal}
  {Journal of Chemical Physics}\ }\textbf {\bibinfo {volume} {144}} (\bibinfo
  {year} {2016})}\BibitemShut {NoStop}%
\bibitem [{\citenamefont {Sato}\ \emph {et~al.}(2016)\citenamefont {Sato},
  \citenamefont {Suzuki}, \citenamefont {Suzuki},\ and\ \citenamefont
  {Adachi}}]{sato2016}%
  \BibitemOpen
  \bibfield  {author} {\bibinfo {author} {\bibfnamefont {M.}~\bibnamefont
  {Sato}}, \bibinfo {author} {\bibfnamefont {Y.}~\bibnamefont {Suzuki}},
  \bibinfo {author} {\bibfnamefont {T.}~\bibnamefont {Suzuki}}, \ and\ \bibinfo
  {author} {\bibfnamefont {S.}~\bibnamefont {Adachi}},\ }\href@noop {}
  {\bibfield  {journal} {\bibinfo  {journal} {Applied Physics Express}\
  }\textbf {\bibinfo {volume} {9}},\ \bibinfo {pages} {022401} (\bibinfo {year}
  {2016})}\BibitemShut {NoStop}%
\bibitem [{\citenamefont {Johnsson}\ \emph {et~al.}(2008)\citenamefont
  {Johnsson}, \citenamefont {Siu}, \citenamefont {Gijsbertsen}, \citenamefont
  {Verhoeven}, \citenamefont {Meijer}, \citenamefont {Van Der~Zande},\ and\
  \citenamefont {Vrakking}}]{johnsson2008}%
  \BibitemOpen
  \bibfield  {author} {\bibinfo {author} {\bibfnamefont {P.}~\bibnamefont
  {Johnsson}}, \bibinfo {author} {\bibfnamefont {W.}~\bibnamefont {Siu}},
  \bibinfo {author} {\bibfnamefont {A.}~\bibnamefont {Gijsbertsen}}, \bibinfo
  {author} {\bibfnamefont {J.}~\bibnamefont {Verhoeven}}, \bibinfo {author}
  {\bibfnamefont {A.}~\bibnamefont {Meijer}}, \bibinfo {author} {\bibfnamefont
  {W.}~\bibnamefont {Van Der~Zande}}, \ and\ \bibinfo {author} {\bibfnamefont
  {M.}~\bibnamefont {Vrakking}},\ }\href@noop {} {\bibfield  {journal}
  {\bibinfo  {journal} {Journal of Modern Optics}\ }\textbf {\bibinfo {volume}
  {55}},\ \bibinfo {pages} {2693} (\bibinfo {year} {2008})}\BibitemShut
  {NoStop}%
\bibitem [{\citenamefont {O'keeffe}\ \emph {et~al.}(2011)\citenamefont
  {O'keeffe}, \citenamefont {Bolognesi}, \citenamefont {Coreno}, \citenamefont
  {Moise}, \citenamefont {Richter}, \citenamefont {Cautero}, \citenamefont
  {Stebel}, \citenamefont {Sergo}, \citenamefont {Pravica}, \citenamefont
  {Ovcharenko} \emph {et~al.}}]{o2011}%
  \BibitemOpen
  \bibfield  {author} {\bibinfo {author} {\bibfnamefont {P.}~\bibnamefont
  {O'keeffe}}, \bibinfo {author} {\bibfnamefont {P.}~\bibnamefont {Bolognesi}},
  \bibinfo {author} {\bibfnamefont {M.}~\bibnamefont {Coreno}}, \bibinfo
  {author} {\bibfnamefont {A.}~\bibnamefont {Moise}}, \bibinfo {author}
  {\bibfnamefont {R.}~\bibnamefont {Richter}}, \bibinfo {author} {\bibfnamefont
  {G.}~\bibnamefont {Cautero}}, \bibinfo {author} {\bibfnamefont
  {L.}~\bibnamefont {Stebel}}, \bibinfo {author} {\bibfnamefont
  {R.}~\bibnamefont {Sergo}}, \bibinfo {author} {\bibfnamefont
  {L.}~\bibnamefont {Pravica}}, \bibinfo {author} {\bibfnamefont
  {Y.}~\bibnamefont {Ovcharenko}},  \emph {et~al.},\ }\href@noop {} {\bibfield
  {journal} {\bibinfo  {journal} {Review of Scientific Instruments}\ }\textbf
  {\bibinfo {volume} {82}},\ \bibinfo {pages} {033109} (\bibinfo {year}
  {2011})}\BibitemShut {NoStop}%
\bibitem [{\citenamefont {Dribinski}\ \emph {et~al.}(2002)\citenamefont
  {Dribinski}, \citenamefont {Ossadtchi}, \citenamefont {Mandelshtam},\ and\
  \citenamefont {Reisler}}]{dribinski2002}%
  \BibitemOpen
  \bibfield  {author} {\bibinfo {author} {\bibfnamefont {V.}~\bibnamefont
  {Dribinski}}, \bibinfo {author} {\bibfnamefont {A.}~\bibnamefont
  {Ossadtchi}}, \bibinfo {author} {\bibfnamefont {V.~A.}\ \bibnamefont
  {Mandelshtam}}, \ and\ \bibinfo {author} {\bibfnamefont {H.}~\bibnamefont
  {Reisler}},\ }\href@noop {} {\bibfield  {journal} {\bibinfo  {journal}
  {Review of Scientific Instruments}\ }\textbf {\bibinfo {volume} {73}},\
  \bibinfo {pages} {2634} (\bibinfo {year} {2002})}\BibitemShut {NoStop}%
\bibitem [{\citenamefont {Manzhos}\ and\ \citenamefont
  {Loock}(2003)}]{manzhos2003}%
  \BibitemOpen
  \bibfield  {author} {\bibinfo {author} {\bibfnamefont {S.}~\bibnamefont
  {Manzhos}}\ and\ \bibinfo {author} {\bibfnamefont {H.-P.}\ \bibnamefont
  {Loock}},\ }\href@noop {} {\bibfield  {journal} {\bibinfo  {journal}
  {Computer physics communications}\ }\textbf {\bibinfo {volume} {154}},\
  \bibinfo {pages} {76} (\bibinfo {year} {2003})}\BibitemShut {NoStop}%
\bibitem [{\citenamefont {Garcia}, \citenamefont {Nahon},\ and\ \citenamefont
  {Powis}(2004)}]{garcia2004}%
  \BibitemOpen
  \bibfield  {author} {\bibinfo {author} {\bibfnamefont {G.~A.}\ \bibnamefont
  {Garcia}}, \bibinfo {author} {\bibfnamefont {L.}~\bibnamefont {Nahon}}, \
  and\ \bibinfo {author} {\bibfnamefont {I.}~\bibnamefont {Powis}},\
  }\href@noop {} {\bibfield  {journal} {\bibinfo  {journal} {Review of
  Scientific Instruments}\ }\textbf {\bibinfo {volume} {75}},\ \bibinfo {pages}
  {4989} (\bibinfo {year} {2004})}\BibitemShut {NoStop}%
\bibitem [{\citenamefont {Roberts}\ \emph {et~al.}(2009)\citenamefont
  {Roberts}, \citenamefont {Nixon}, \citenamefont {Lecointre}, \citenamefont
  {Wrede},\ and\ \citenamefont {Verlet}}]{roberts2009}%
  \BibitemOpen
  \bibfield  {author} {\bibinfo {author} {\bibfnamefont {G.}~\bibnamefont
  {Roberts}}, \bibinfo {author} {\bibfnamefont {J.}~\bibnamefont {Nixon}},
  \bibinfo {author} {\bibfnamefont {J.}~\bibnamefont {Lecointre}}, \bibinfo
  {author} {\bibfnamefont {E.}~\bibnamefont {Wrede}}, \ and\ \bibinfo {author}
  {\bibfnamefont {J.}~\bibnamefont {Verlet}},\ }\href@noop {} {\bibfield
  {journal} {\bibinfo  {journal} {Review of Scientific Instruments}\ }\textbf
  {\bibinfo {volume} {80}},\ \bibinfo {pages} {053104} (\bibinfo {year}
  {2009})}\BibitemShut {NoStop}%
\bibitem [{\citenamefont {Gebhardt}\ \emph {et~al.}(2001)\citenamefont
  {Gebhardt}, \citenamefont {Rakitzis}, \citenamefont {Samartzis},
  \citenamefont {Ladopoulos},\ and\ \citenamefont
  {Kitsopoulos}}]{gebhardt2001}%
  \BibitemOpen
  \bibfield  {author} {\bibinfo {author} {\bibfnamefont {C.~R.}\ \bibnamefont
  {Gebhardt}}, \bibinfo {author} {\bibfnamefont {T.~P.}\ \bibnamefont
  {Rakitzis}}, \bibinfo {author} {\bibfnamefont {P.~C.}\ \bibnamefont
  {Samartzis}}, \bibinfo {author} {\bibfnamefont {V.}~\bibnamefont
  {Ladopoulos}}, \ and\ \bibinfo {author} {\bibfnamefont {T.~N.}\ \bibnamefont
  {Kitsopoulos}},\ }\href@noop {} {\bibfield  {journal} {\bibinfo  {journal}
  {Review of Scientific Instruments}\ }\textbf {\bibinfo {volume} {72}},\
  \bibinfo {pages} {3848} (\bibinfo {year} {2001})}\BibitemShut {NoStop}%
\bibitem [{\citenamefont {Townsend}, \citenamefont {Minitti},\ and\
  \citenamefont {Suits}(2003)}]{townsend2003}%
  \BibitemOpen
  \bibfield  {author} {\bibinfo {author} {\bibfnamefont {D.}~\bibnamefont
  {Townsend}}, \bibinfo {author} {\bibfnamefont {M.~P.}\ \bibnamefont
  {Minitti}}, \ and\ \bibinfo {author} {\bibfnamefont {A.~G.}\ \bibnamefont
  {Suits}},\ }\href@noop {} {\bibfield  {journal} {\bibinfo  {journal} {Review
  of scientific instruments}\ }\textbf {\bibinfo {volume} {74}},\ \bibinfo
  {pages} {2530} (\bibinfo {year} {2003})}\BibitemShut {NoStop}%
\bibitem [{\citenamefont {Lee}\ \emph {et~al.}(2014{\natexlab{a}})\citenamefont
  {Lee}, \citenamefont {Lin}, \citenamefont {Lingenfelter}, \citenamefont
  {Fan}, \citenamefont {Winney},\ and\ \citenamefont
  {Li}}]{lee2014communication}%
  \BibitemOpen
  \bibfield  {author} {\bibinfo {author} {\bibfnamefont {S.~K.}\ \bibnamefont
  {Lee}}, \bibinfo {author} {\bibfnamefont {Y.~F.}\ \bibnamefont {Lin}},
  \bibinfo {author} {\bibfnamefont {S.}~\bibnamefont {Lingenfelter}}, \bibinfo
  {author} {\bibfnamefont {L.}~\bibnamefont {Fan}}, \bibinfo {author}
  {\bibfnamefont {A.~H.}\ \bibnamefont {Winney}}, \ and\ \bibinfo {author}
  {\bibfnamefont {W.}~\bibnamefont {Li}},\ }\href@noop {} {\bibfield  {journal}
  {\bibinfo  {journal} {The Journal of chemical physics}\ }\textbf {\bibinfo
  {volume} {141}},\ \bibinfo {pages} {221101} (\bibinfo {year}
  {2014}{\natexlab{a}})}\BibitemShut {NoStop}%
\bibitem [{\citenamefont {Rolles}\ \emph {et~al.}(2007)\citenamefont {Rolles},
  \citenamefont {Pe{\v{s}}i{\'c}}, \citenamefont {Perri}, \citenamefont
  {Bilodeau}, \citenamefont {Ackerman}, \citenamefont {Rude}, \citenamefont
  {Kilcoyne}, \citenamefont {Bozek},\ and\ \citenamefont
  {Berrah}}]{rolles2007}%
  \BibitemOpen
  \bibfield  {author} {\bibinfo {author} {\bibfnamefont {D.}~\bibnamefont
  {Rolles}}, \bibinfo {author} {\bibfnamefont {Z.}~\bibnamefont
  {Pe{\v{s}}i{\'c}}}, \bibinfo {author} {\bibfnamefont {M.}~\bibnamefont
  {Perri}}, \bibinfo {author} {\bibfnamefont {R.}~\bibnamefont {Bilodeau}},
  \bibinfo {author} {\bibfnamefont {G.}~\bibnamefont {Ackerman}}, \bibinfo
  {author} {\bibfnamefont {B.}~\bibnamefont {Rude}}, \bibinfo {author}
  {\bibfnamefont {A.}~\bibnamefont {Kilcoyne}}, \bibinfo {author}
  {\bibfnamefont {J.}~\bibnamefont {Bozek}}, \ and\ \bibinfo {author}
  {\bibfnamefont {N.}~\bibnamefont {Berrah}},\ }\href@noop {} {\bibfield
  {journal} {\bibinfo  {journal} {Nuclear Instruments and Methods in Physics
  Research Section B: Beam Interactions with Materials and Atoms}\ }\textbf
  {\bibinfo {volume} {261}},\ \bibinfo {pages} {170} (\bibinfo {year}
  {2007})}\BibitemShut {NoStop}%
\bibitem [{\citenamefont {Lee}\ \emph {et~al.}(2014{\natexlab{b}})\citenamefont
  {Lee}, \citenamefont {Cudry}, \citenamefont {Lin}, \citenamefont
  {Lingenfelter}, \citenamefont {Winney}, \citenamefont {Fan},\ and\
  \citenamefont {Li}}]{lee2014}%
  \BibitemOpen
  \bibfield  {author} {\bibinfo {author} {\bibfnamefont {S.~K.}\ \bibnamefont
  {Lee}}, \bibinfo {author} {\bibfnamefont {F.}~\bibnamefont {Cudry}}, \bibinfo
  {author} {\bibfnamefont {Y.~F.}\ \bibnamefont {Lin}}, \bibinfo {author}
  {\bibfnamefont {S.}~\bibnamefont {Lingenfelter}}, \bibinfo {author}
  {\bibfnamefont {A.~H.}\ \bibnamefont {Winney}}, \bibinfo {author}
  {\bibfnamefont {L.}~\bibnamefont {Fan}}, \ and\ \bibinfo {author}
  {\bibfnamefont {W.}~\bibnamefont {Li}},\ }\href@noop {} {\bibfield  {journal}
  {\bibinfo  {journal} {Review of Scientific Instruments}\ }\textbf {\bibinfo
  {volume} {85}},\ \bibinfo {pages} {123303} (\bibinfo {year}
  {2014}{\natexlab{b}})}\BibitemShut {NoStop}%
\bibitem [{\citenamefont {Macklin}, \citenamefont {Kmetec},\ and\ \citenamefont
  {Gordon~III}(1993)}]{macklin1993}%
  \BibitemOpen
  \bibfield  {author} {\bibinfo {author} {\bibfnamefont {J.}~\bibnamefont
  {Macklin}}, \bibinfo {author} {\bibfnamefont {J.}~\bibnamefont {Kmetec}}, \
  and\ \bibinfo {author} {\bibfnamefont {C.}~\bibnamefont {Gordon~III}},\
  }\href@noop {} {\bibfield  {journal} {\bibinfo  {journal} {Physical Review
  Letters}\ }\textbf {\bibinfo {volume} {70}},\ \bibinfo {pages} {766}
  (\bibinfo {year} {1993})}\BibitemShut {NoStop}%
\bibitem [{\citenamefont {Krause}, \citenamefont {Schafer},\ and\ \citenamefont
  {Kulander}(1992)}]{krause1992}%
  \BibitemOpen
  \bibfield  {author} {\bibinfo {author} {\bibfnamefont {J.~L.}\ \bibnamefont
  {Krause}}, \bibinfo {author} {\bibfnamefont {K.~J.}\ \bibnamefont {Schafer}},
  \ and\ \bibinfo {author} {\bibfnamefont {K.~C.}\ \bibnamefont {Kulander}},\
  }\href@noop {} {\bibfield  {journal} {\bibinfo  {journal} {Physical Review
  Letters}\ }\textbf {\bibinfo {volume} {68}},\ \bibinfo {pages} {3535}
  (\bibinfo {year} {1992})}\BibitemShut {NoStop}%
\bibitem [{\citenamefont {Corkum}(1993)}]{corkum1993}%
  \BibitemOpen
  \bibfield  {author} {\bibinfo {author} {\bibfnamefont {P.~B.}\ \bibnamefont
  {Corkum}},\ }\href@noop {} {\bibfield  {journal} {\bibinfo  {journal}
  {Physical Review Letters}\ }\textbf {\bibinfo {volume} {71}},\ \bibinfo
  {pages} {1994} (\bibinfo {year} {1993})}\BibitemShut {NoStop}%
\bibitem [{\citenamefont {Shivaram}\ \emph {et~al.}(2010)\citenamefont
  {Shivaram}, \citenamefont {Roberts}, \citenamefont {Xu},\ and\ \citenamefont
  {Sandhu}}]{Shivaram2010}%
  \BibitemOpen
  \bibfield  {author} {\bibinfo {author} {\bibfnamefont {N.}~\bibnamefont
  {Shivaram}}, \bibinfo {author} {\bibfnamefont {A.}~\bibnamefont {Roberts}},
  \bibinfo {author} {\bibfnamefont {L.}~\bibnamefont {Xu}}, \ and\ \bibinfo
  {author} {\bibfnamefont {A.}~\bibnamefont {Sandhu}},\ }\href {<Go to
  ISI>://WOS:000283048100002} {\bibfield  {journal} {\bibinfo  {journal}
  {Optics Letters}\ }\textbf {\bibinfo {volume} {35}},\ \bibinfo {pages} {3312}
  (\bibinfo {year} {2010})}\BibitemShut {NoStop}%
\bibitem [{\citenamefont {Stei}\ \emph {et~al.}(2013)\citenamefont {Stei},
  \citenamefont {von Vangerow}, \citenamefont {Otto}, \citenamefont {Kelkar},
  \citenamefont {Carrascosa}, \citenamefont {Best},\ and\ \citenamefont
  {Wester}}]{Stei2013}%
  \BibitemOpen
  \bibfield  {author} {\bibinfo {author} {\bibfnamefont {M.}~\bibnamefont
  {Stei}}, \bibinfo {author} {\bibfnamefont {J.}~\bibnamefont {von Vangerow}},
  \bibinfo {author} {\bibfnamefont {R.}~\bibnamefont {Otto}}, \bibinfo {author}
  {\bibfnamefont {A.~H.}\ \bibnamefont {Kelkar}}, \bibinfo {author}
  {\bibfnamefont {E.}~\bibnamefont {Carrascosa}}, \bibinfo {author}
  {\bibfnamefont {T.}~\bibnamefont {Best}}, \ and\ \bibinfo {author}
  {\bibfnamefont {R.}~\bibnamefont {Wester}},\ }\href {<Go to
  ISI>://WOS:000320138900014} {\bibfield  {journal} {\bibinfo  {journal}
  {Journal of Chemical Physics}\ }\textbf {\bibinfo {volume} {138}} (\bibinfo
  {year} {2013})}\BibitemShut {NoStop}%
\end{thebibliography}%

% Optionally you can create manual references
%\begin{thebibliography}
%\bibitem{lastie2005}	X. Y. Lastie, A. K. Grim, S. V. Moro, D. Jia, and A. A. Firs, {\it Nature} {\bf 400}, 197-200 (2005).
%\end{thebibliography}
% Produces the bibliography via BibTeX

\end{document}